\documentclass[a4paper,aps,twoside,twocolumn]{revtex4}

\usepackage{epsfig}
\usepackage{amsmath,amssymb,color}
\usepackage{footnote}
\usepackage[cp1251]{inputenc}
\usepackage[english]{babel}

\parskip=\medskipamount

\newcommand{\eq}[1]{(\ref{#1})}
\newcommand{\fig}[1]{Fig.\ref{#1}}

\newcommand{\be}{\begin{equation}}
\newcommand{\ee}{\end{equation}}

\newcommand{\barr}{\begin{array}}
\newcommand{\earr}{\end{array}}

\newcommand{\beqn}{\begin{eqnarray}}
\newcommand{\eeqn}{\end{eqnarray}}

\newcommand{\bs}{\begin{subequations}}
\newcommand{\es}{\end{subequations}}

\newcommand{\bw}{\begin{widetext}}
\newcommand{\ew}{\end{widetext}}

\begin{document}

\title{KPZ scaling in topological mixing}

\author{M. Beltr\'an del R\'\i o$^{1,2}$, S. Nechaev$^{2,3,4}$, M. Taran$^{5}$}

\affiliation{$^1$Instituto de F\'\i sica, Universidad Nacional Aunt\'onoma de M\'exico,
Departamento de Sistemas Complejos \\ $^2$LPTMS, Universit\'e Paris Sud, 91405 Orsay Cedex, France \\
$^3$P.N. Lebedev Physical Institute of the Russian Academy of Sciences, 119991, Moscow, Russia \\
$^4$J.-V. Poncelet Labotatory, Independent University, 119002, Moscow, Russia \\
$^5$State Scientific Center TRINITI, 142190, Troitsk, Russia}

\date{\today}

\begin{abstract}

In the spirit of recent works on topological chaos generated by sequential rotation of infinitely
thin stirrers placed in a viscous liquid, we consider the statistical properties of braiding
exponent which quantitatively characterizes the chaotic behavior of advected particles in
two--dimensional flows. We pay a special attention to the random stirring protocol and study the
time--dependent behavior of the variance of the braiding exponent. We show that this behavior
belongs to the Kardar--Parisi--Zhang universality class typical for models of nonstationary growth.
Using the matrix (Magnus) representation of the braid group generators, we relate the random
stirring protocol with the growth of random heap generated by a ballistic deposition.

\bigskip

\noindent PACS numbers:

\end{abstract}

\maketitle

Finding the ways of efficient mixing in liquids is one of the very popular subjects in physics of
fluids because of numerous important technological applications ranging from the industrial
production of blends and alloys to the percolation of turbulent flows of viscous liquids through
porous media. In connection with that the experimental and theoretical study of the chaotic motion
of passively advected particles in two--dimensional flows in presence of moving rods ("stirrers")
become very popular during the last decade after the key work \cite{boyland1} generalized and
developed later in \cite{boyland2,finn,vikh1,vikh2,thiff1,chen,thiff2}. The most attention in these
works is paid to the following question: how should the stirrers (playing the role of topological
obstacles in the liquid) move to produce the strong chaos in passively advected particles in
viscous liquid. This question is known as finding the best "stirring protocol". The presence of the
strong chaotic behavior due to motion of stirrers is guaranteed by topological arguments based on
the Thurston--Nielsen theory \cite{thurston,casson} which postulates the existence of regions in
the flow with pseudo--Anosov dynamics when every point of the flow demonstrates the exponential
stretching with positive Lyapunov exponent.

On the basis of topological arguments it has been pointed out in \cite{boyland1} that the chaotic
behavior in the flow of advected particles can be quantitatively characterized by measuring the
Lyapunov exponents of the product of matrices representing the sequential permutation of
neighboring stirrers in space. To be more precise, the following "discretized" model has been
considered (see, for example, \cite{bouland1,tang}). Suppose that the viscous liquid passes through
the comb of stirrers and in the same short intervals of time two neighboring in the comb stirrers
permute carrying along some part of a liquid including a trace of advected particle. In the most
transparent way such a model has been formulated in \cite{thiff1}. In this work the liquid passed
through the comb of $n$ aligned infinitely thin rods. The permutations of neighboring rods leads to
the entanglement of the trace of advected particle and can be characterized by a generator of the
{\em braid group} $B_{n+1}$ (the definition is given below). Thus, the time--ordered product of $T$
sequential permutations, $W_T=:\hspace{-3pt}\prod_{t=1}^{T}\sigma_{i_t}\hspace{-3pt}:$ is a word
written in terms of elementary operations -- permutations of neighboring stirrers. Each generator
of the braid group, $\sigma_i$ ($1 \le i \le n$) representing the permutation of stirrers $i$ and
$i+1$ in the comb can be written in the matrix (Burau) representation  (see, for example,
\cite{birman}). Thus, the word $W_T$ is an $n\times n$ square matrix $\hat{W}_T$ containing the
full information about the "stirring protocol". Computing the Lyapunov exponents
$\lambda\{\hat{W}_T\}$ for different sequences of permutations and looking at the dependence
$\lambda(T)$ one can conclude about the properties of the mixing regime.

In this letter we formulate the mixing problem in a slightly different manner paying the most
attention to its statistical aspects. Briefly, the outline of our consideration is as follows. We
suppose that the matrix $\hat{W}_T$ is the time--ordered product of $T$ generators of the braid
group taken in a {\it random} order with the uniform distribution over the set generators
$\{\sigma_1,\sigma_2,..., \sigma_n\}$. In terms of the works \cite{boyland1,thiff1,thiff2} this
procedure of mixing can be regarded as a {\em random stirring protocol}. For each generated
sequence $\hat{W}_T$ (for fixed $T$) we compute the associated Lyapunov exponent,
$\lambda\{\hat{W}_T\}$. Averaging $\lambda\{\hat{W}_T\}$ over the ensemble of all possible
sequences $W_T$ we find the mean value, $\overline{\lambda}(T)$; also we compute the variance,
$u^2(T)= \overline{\left(\lambda\{W_T\}-\overline{\lambda}(T)\right)^2}$. Our main result consists
in the following. We show that in the broad interval of values $T$ and $n$, the variance $u(T,n)$
demonstrates the Kardar--Parisi--Zhang (KPZ) scaling (see, for review \cite{kpz}), i.e.
$u(T,n)=n^{1/2}g(T/n^{3/2})$ with $g(u)\sim u^{1/3}$ for $u\ll 1$ and $g(u)\sim {\rm const}$ for
$u\gg 1$. Such a behavior is typical for many non-stationary growth problems. It is believed now
that KPZ universality class for {\em correlated} stochastic processes is as typical as the Gaussian
statistics for uncorrelated ones.

\begin{figure}[ht]
\epsfig{file=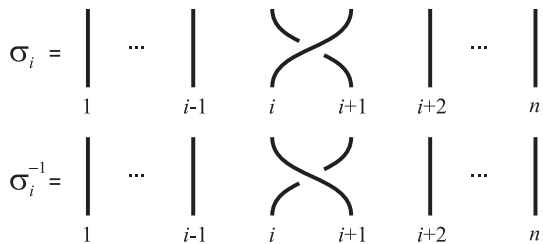,width=7cm}
\caption{Braid group generator $\sigma_i$ and $\sigma_i^{-1}$.}
\label{fig:1}
\end{figure}

To make the content of this letter as selfconsistent as possible, it is instructive to introduce
the basic definitions and precisely formulate the model under consideration. We begin with the
definition of the braid group, $B_n$. The group $B_{n+1}$ has $n$ generators $\sigma_1,...,
\sigma_n$ (and their inverses) with the commutation relations:
\be
\left\{
\begin{array}{cc}
\sigma_i \sigma_{i+1} \sigma_i = \sigma_{i+1} \sigma_i \sigma_{i+1} & \quad 1\le i \le n \medskip \\
\sigma_i \sigma_k = \sigma_k \sigma_i & \quad  |i-k|\ge 2
\end{array} \right.
\label{eq:1}
\ee
Any arbitrary word written in terms of letters--generators from the set $\{\sigma_1,...
\sigma_n,\sigma_1^{-1},..., \sigma_n^{-1}\}$---gives a particular {\it braid}. The graphic
representation of the braid generators is shown in \fig{fig:1}a. The length of the braid is the
total number of used letters. Diagrammatically the braid can be represented by a set of crossing
strings going from the top to the bottom after subsequent gluing the braid generators.

The generators of the groups $B_n$ admit the matrix representations. For our purposes it is more
convenient to use for the braid group generator $\sigma_i$ the Marnus representation instead of the
Bureau one \cite{birman}. Thus, we have for $\sigma_i$:
\be
\sigma_i=\left(\begin{array}{ccccccccc}
1     & 0    &\cdots             &      &      \\
0     &\ddots&                   &      &      \\
\vdots&      &\framebox{$A$}&      &\vdots\\
&     &                          &\ddots& 0    \\
&     &\cdots                    & 0    & 1
\end{array}\right)\leftarrow \mbox{row $i$}; \;
A=\left(\begin{array}{ccc} 1 & 0 & 0 \\ u & -u & 1 \\ 0 & 0 & 1 \end{array} \right)
\label{eq:3}
\ee
For different $i$'s ($2\le i \le n-1$) the $3\times 3$ blocks $A$ in \eq{eq:3} "slides" along the
diagonal, while for $i=1$ and $i=n$ the corner blocks are correspondingly $A'$ and $A''$, where
\be
A'=\left(\begin{array}{cc} -u & 1 \\ 0 & 1 \end{array}\right); \quad A''=\left(\begin{array}{cc} 1 & 0 \\
u & -u
\end{array} \right)
\label{eq:3a}
\ee
Note, that the matrices $A'$ and $A''$ are the generators of the simplest nontrivial group $B_3$.
It is known \cite{birman} that $B_3$ for $u=-1$ coincides with the modular group $PSL(2,Z)$, i.e.
the group $B_3$ is the central extension of $PSL(2,Z)$. To speak rigorously, from this point of
view, the matrix representation of $B_3$ used in \cite{boyland1} and in the consecutive works deals
with the group $PSL(2,Z)$, but not with $B_3$.

Now we are in position to precisely formulate the model under consideration and the main question
of our interest. In what follows we consider for simplicity the semigroup $B_{n+1}^+$, i.e. we
construct the words of letters (generators) from the set $\{\sigma_1,,,, \sigma_n\}$ only (without
the inverses). The extension of our consideration on the full group and related difficulties are
discussed briefly at the end of this letter.

Define the {\em random walk} on a semigroup $B_{n+1}^+$. Let $\mu=\frac{1}{n}$ be the uniform
measure on the sets of generators of $B_{n+1}^+$. Consider the (right--hand side) random walk on
$B_{n+1}^+$ with fixed framings, i.e. regard the Markov chain with the following transition
probabilities: the word $W$ transforms into $W\,\sigma$, where $\sigma$ is one of the generators
from the set $\{\sigma_1,...,\sigma_n\}$ taken with the uniform probability $\mu$. For any
particular $T$--step random walk on the semigroup $B_n^+$ we construct the word (the matrix)
$\hat{W}_T\{B_{n+1}^+\}$ and compute the associated largest eigenvalue $\Lambda_T(u)$ which in the
terminology of the work \cite{thiff1} is called the {\em braiding factor}. Then we extract the {\em
maximal} Lyapunov ("braiding") exponent $\lambda_T$ by taking the limit
\be
\lambda_T =  \lim_{u\to\infty}\frac{\ln |\Lambda_T(u)|}{\ln u}
\label{eq:5}
\ee

Note the difference between \eq{eq:5} and the definition of the braiding factor in \cite{thiff1}.
Actually, to find the maximal Lyapunov exponent we take the limit with respect to the dummy
variable $u$ which enters in the definition of the matrix representation of the braid group
generator \eq{eq:3}, while in \cite{thiff1} the limit is taken with respect to the "time" i.e. the
length of the word, $T$. Below we present the arguments which justify our definition making it
topologically and physically transparent.

Since the word $W_T$ is random, the associated Lyapunov exponent $\lambda_T$ is also a random
value. Now we proceed as it has been stated above. We average $\lambda\{\hat{W}_T\}$ over the
ensemble of all possible sequences $W_T$ and find the mean value, $\overline{\lambda}(T)$. Then we
compute the variance, $u^2(T)= \overline{\left(\lambda\{W_T\}-\overline{\lambda}(T)\right)^2}$ and
consider its dependence against $T$. The function $u(T)$ is the central result of our letter.

The matrix representation \eq{eq:3} introduced above permits us to link the topological problems
with the "Tetris--like" models models of {\em ballistic deposition}. Here the notion of the random
walk on the semigroup $B_n^+$ becomes very useful. On one hand, we see from \fig{fig:1} that the
generator $\sigma_i$ has the clear topological meaning, thus the information about the degree of
entanglement of threads (i.e. about the braiding factor) is encoded in the word $W_T$. On the other
hand, the word written in terms of generators of $B_n^+$ can be interpreted as a {\em heap} (or
pile) obtained by a 1+1 sequential ballistic deposition process in a bounding box. The matrix
representation \eq{eq:3} allows to determine the corresponding height profile. Let us discuss this
connection in more details.

\begin{figure}[ht]
\epsfig{file=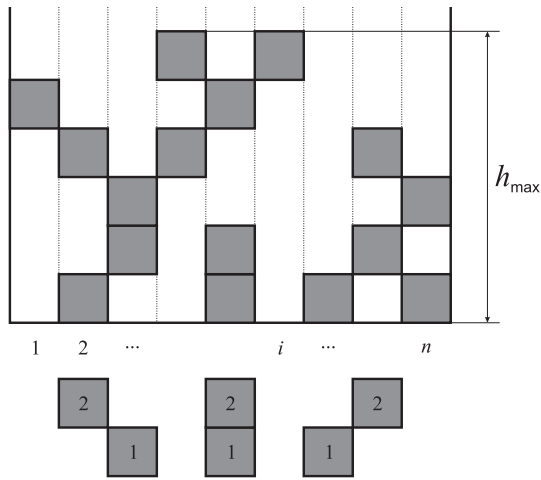,width=7cm}
\caption{The heap created by sequential deposition of blocks. The interactions of two sequential
blocks (1st and 2nd) are shown.}
\label{fig:2}
\end{figure}

A standard one--dimensional ballistic deposition model with next--nearest--neighboring (NNN)
interactions is formulated as follows (for more details, see Refs.\cite{scaling1,scaling2,
scaling3}). Consider a box divided in $n$ columns (of unit width each) enumerated by an index $i$
($i=1,2,...,n$). The free conditions are assumed for left and right boundaries. At the initial time
moment, ($t=0$), the system is empty. Then, at each tick of the clock, $t=1,2,...,T$,  we deposit
an elementary cell ("particle") of unit height and width in a randomly chosen column, $i$. Suppose
that the distribution on the set of columns is uniform. Define the height, $h_i(t)$, in the column
$i$ at time moment $t$. Assume now, as it is depicted in \fig{fig:2}, that the cells in the
nearest--neighboring columns interact in such a way that they can only touch each other by corners,
but never by their vertical sides. This implies that after having deposited a particle to the
column $i$, the height of this column is modified according to the following rule:
\be
h_i(t+1)=\max[h_{i-1}(t),\, h_i(t),\, h_{i+1}(t)]+1
\label{eq:6}
\ee
If at the time moment $t$ nothing is added to the column $i$, its height remains unchanged:
$h_i(t+1)= h_i(t)$. A set of deposited particles forms a heap (a pile) as shown in \fig{fig:2}.

The height $h_i(t)$ ($1\le i \le n$) in the column $i$ at time $t$ can be easily computed using the
matrix representation similar to \eq{eq:3}. The deposition event in the column $i$ means the
application of the generator $g_i$ with the following matrix representation:
\be
g_i=\left(\begin{array}{ccccccccc}
1     & 0    &\cdots             &      &      \\
0     &\ddots&                   &      &      \\
\vdots&      &\framebox{$B$}&      &\vdots\\
&     &                          &\ddots& 0    \\
&     &\cdots                    & 0    & 1
\end{array}\right)\leftarrow \mbox{row $i$}; \;
B=\left(\begin{array}{ccc} 1 & 0 & 0 \\ u & u & u \\ 0 & 0 & 1 \end{array} \right)
\label{eq:4}
\ee
Graphically the deposition process is associated with a sequence of "dropping events" of elementary
blocks--generators $g_i$ ($1\le i\le n$). Since we rise a heap by sequential multiplication of
matrices \eq{eq:4}, we arrive at the matrix $\hat{U}_T(n,u)=:\hspace{-3pt} \prod_{t=1}^{T}
g_{i_t}\hspace{-3pt}:$ ($1\le t\le T$). Each matrix element of $\hat{U}_T(n,u)$ is a polynomial of
the variable $u$. Take now a vector column ${\bf a}(t=0)=(a_1,...,a_n)$ where the components of
${\bf a}(t=0)$ are distinct nonzero values. The local heights ${\bf h}(T)=(h_1(T),...,h_n(T))$ we
extract as follows:
\be
{\bf h}(T) = \lim_{u\to\infty} \frac{\ln [\hat{U}_T(n,u)\, {\bf a}(t=0)]}{\ln u}
\label{eq:7}
\ee
Let us demonstrate that the definition of heights in \eq{eq:4}--\eq{eq:7} is consistent with the
updating rules \eq{eq:6} for the NNN--ballistic deposition process. Using \eq{eq:4} we can write
the recursion relation for the dynamics of the vector ${\bf a}(t)$:
\be
{\bf a}(t+1) = g_{i_t} {\bf a}(t); \quad (i_t\in[1,n])
\label{eq:rec1}
\ee
If $i_t=i$, then
\be
a_i(t+1) = u\, a_{i-1}(t) + u\, a_i(t) + u\, a_{i+1}(t)
\label{eq:rec2}
\ee
Supposing that $a_i(t) = u^{h_i(t)}$ and substituting this ansatz into \eq{eq:rec2}, we get:
\be
h_i(t+1) = \frac{e^{(h_{i-1}(t)+1)\ln u } + e^{(h_i(t)+1)\ln u} + e^{(h_{i+1}(t)+1)\ln u}}{\ln u}
\label{eq:rec3}
\ee
In the limit $u\to \infty$ Eq.\eq{eq:rec3} coincides with Eq.\eq{eq:6}.

\begin{figure}[ht]
\epsfig{file=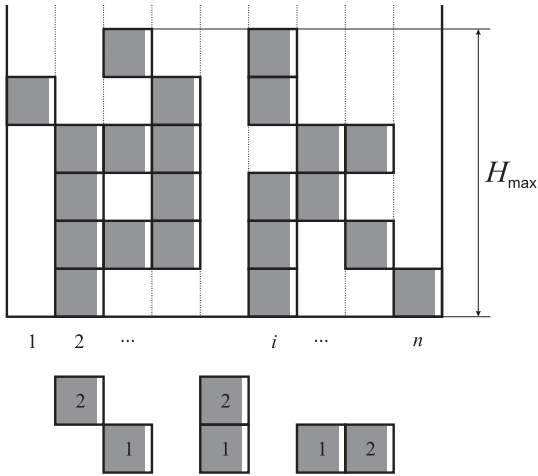,width=7cm}
\caption{The heap created by blocks associated with generators of the semigroup $B_{n+1}^+$. The
"one--sided--sticky" interactions of two sequential blocks (1st and 2nd) are shown.}
\label{fig:3}
\end{figure}

The similarity between the representations \eq{eq:3} and \eq{eq:4} allows us to expect that the
sequential multiplication of braid semigroup generators can also be interpreted a sort of ballistic
deposition process. To see that let us first make the following temporary replacement for the
$3\times 3$ block $A$ in \eq{eq:3}:
\be
A=\left(\begin{array}{ccc} 1 & 0 & 0 \\ u & -u & 1 \\ 0 & 0 & 1 \end{array}\right) \to
\tilde{A}=\left(\begin{array}{ccc} 1 & 0 & 0 \\ u & u & 1 \\ 0 & 0 & 1 \end{array}\right)
\label{eq:8}
\ee
By this exchange we ensure for $\tilde{A}$ the absence of any cancellations of terms of polynomials
in $u$ in course of matrix multiplications since all matrix elements stay positive. We show later
that even for the true block $A$ such cancellations are exponentially rare do not change the
obtained results. Now the multiplication of matrices \eq{eq:3} with the replacement $A\to
\tilde{A}$ has the meaning of the heap construction shown in \fig{fig:3} with slightly modified
local interactions. One can imagine that every elementary block has "sticky" right--hand side as it
is depicted in \fig{fig:3}.

The corresponding height $H_i(t)$ after depositing a particle to the column $i$ at time $t$ is
modified as follows:
\be
H_i(t+1)=\max\{H_{i-1}(t),\, H_i(t)+1,\, H_{i+1}(t)+1\}
\label{eq:9}
\ee
If at the time moment $t$ nothing is added to the column $i$, its height remains unchanged:
$H_i(t+1)= H_i(t)$.

\begin{figure}[ht]
\epsfig{file=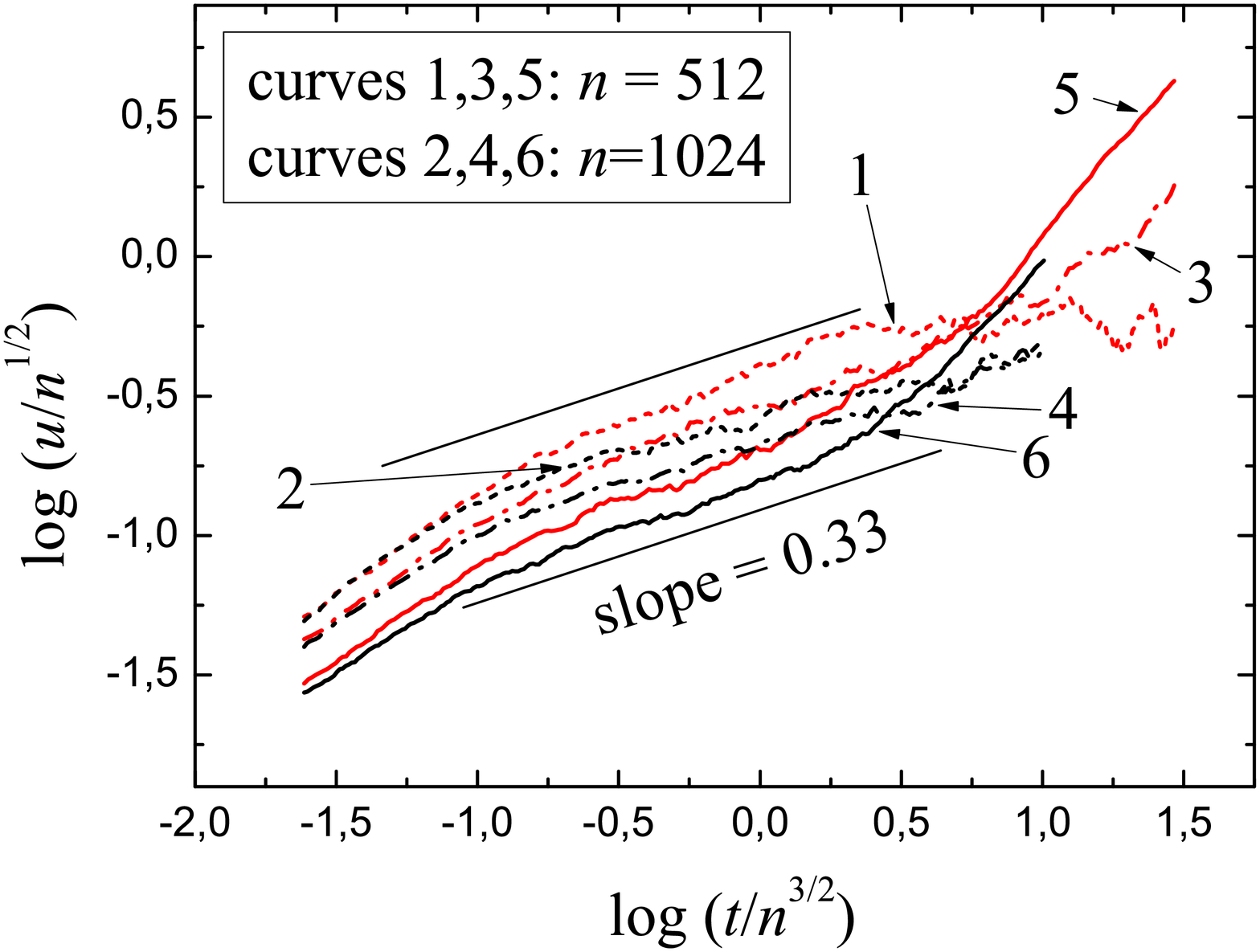,width=7cm}
\caption{KPZ scaling in: the standard NNN--ballistic deposition (curves 1 and 2); the ballistic
deposition of "one--sided--sticky" particles (curves 3 and 4); multiplication of braid semigroup
generators (curves 5 and 6).}
\label{fig:4}
\end{figure}

Our first result concerns the computation of the height profiles for deposition processes depicted
in \fig{fig:2} and \fig{fig:3}. The main quantity of interest is the variance $w_h(T,n)$ defined
for the standard NNN--deposition process as follows:
\be
w_h(T,n)=\frac{1}{n^{1/2}} \left[\sum_{i=1}^{n}\left(\overline{h} - h_i\right)^2 \right]^{1/2};
\quad \overline{h}= \frac{1}{n}\sum_{i=1}^n h_i
\label{eq:10}
\ee
The mean value $\overline{H}$ and the variance $w_H(T,n)$ for the ballistic deposition of
one--sided--sticky particles (see \fig{fig:3}) are defined in the same way. We plot in \fig{fig:4}
the dependence $w_H(t)$ in the coordinates $\left(w/n^{1/2},\; t/n^{3/2}\right)$ implying the known
KPZ scaling $w_H=n^{1/2}f(t/n^{3/2})$ with $f(u)\sim u^{1/3}$ for $u\ll 1$ and $f(u)\sim {\rm
const}$ for $u\gg 1$.

Comparing figures \fig{fig:2} and \fig{fig:3} we can expect that the modifications of local
NNN--interactions would not affect the scaling behavior of the width of the height distribution in
a growing heap defined by \eq{eq:9}. This is actually so and as it can be seen by comparing curves
(1,2) with (3,4) in \fig{fig:4}. Thus, we recover the KPZ scaling for the variance $w_H(t,n)$ (as
the function of "time" $t$.

Now we return to the random multiplication of matrices \eq{eq:3} -- generators  of the braid
semigroup, $W_T=:\hspace{-3pt}\prod_{t=1}^{T}\sigma_{i_t}\hspace{-3pt}:$ and numerically compute
the associated "height", ${\bf Y}=(y_1(T),...y_n(T))$, defined similarly to \eq{eq:7}:
\be
{\bf Y}(T) = \lim_{u\to\infty} \frac{\ln [\hat{W}_T(n,u)\, {\bf a}(t=0)]}{\ln u}
\label{eq:11}
\ee
The interpretation of ${\bf Y}$ as a hight is still valid however some care should be taken to
check the absence of statistically significant amount of cancellations due to the "$-$" sign in the
block $A$ in \eq{eq:3}. The possible cancellations are due to the commensurability effects and are
negligible for generic initial set ${\bf a}(t=0)=(a_1,...,a_n)$ (for instance, if $a_i$, $1\le i
\le n$, are distinct real values from the interval $]0,1[$). The corresponding scaling dependence
is shown in \fig{fig:4} (curves 5 and 6).

\begin{figure}[ht]
\epsfig{file=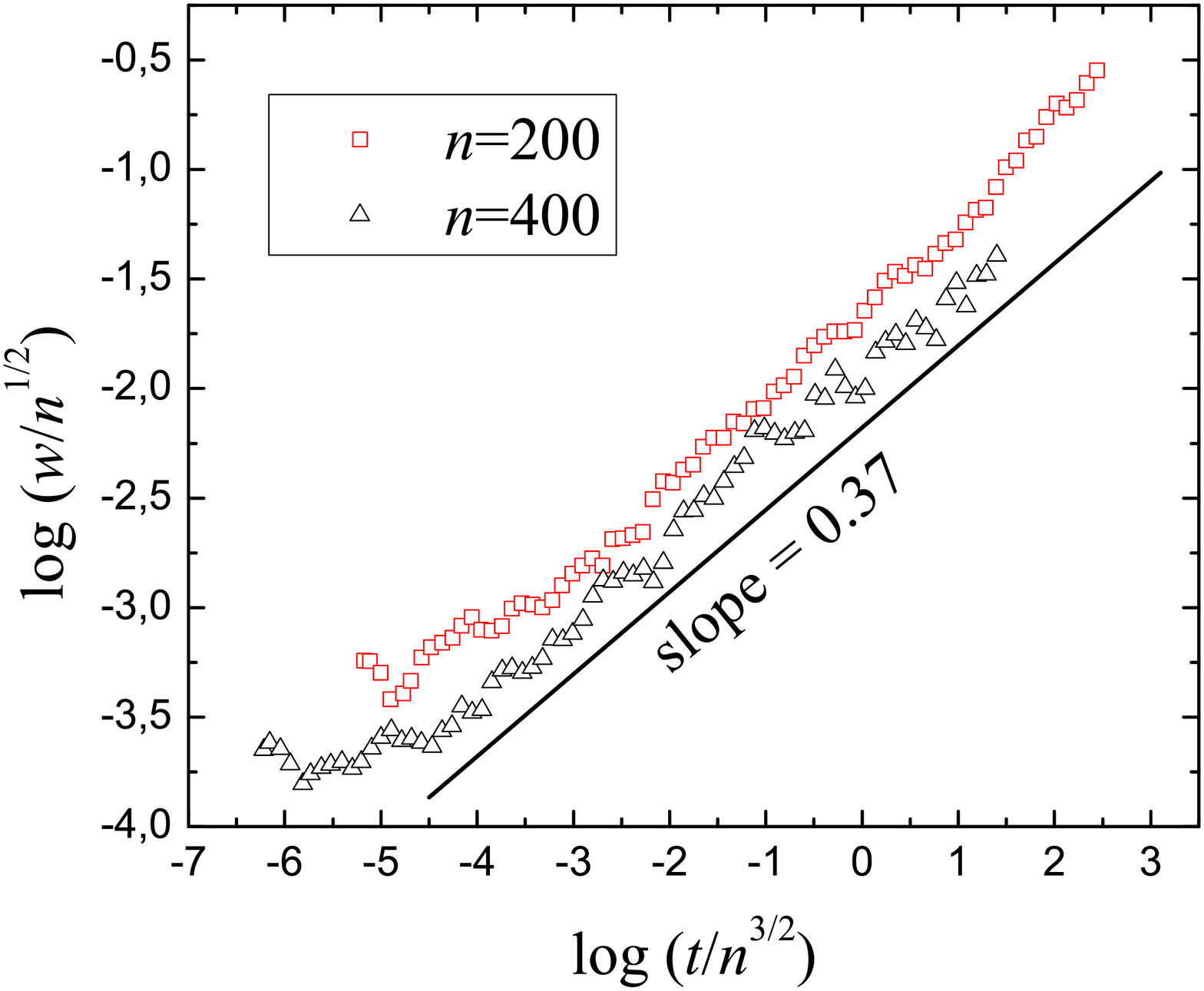,width=7cm}
\caption{KPZ scaling in double logarithmic scale for the variances in the multiplication of matrices
\eq{eq:3} with the "true" block $A$.}
\label{fig:5}
\end{figure}

The obtained KPZ scaling for the "heights" depicted in \fig{fig:4} associated with the random
multiplication of braiding matrices indicates the possibility of similar behavior for the braiding
exponent $\lambda_T$ defined in \eq{eq:5}. To check that we have multiplied sequentially $T$
randomly taken (with uniform distribution $\mu=\frac{1}{n-1}$) matrices---generators of the braid
semigroup $B_n^+$---from the set $\{\sigma_1,...,\sigma_{n-1}\}$ and obtained the associated "word"
in the matrix representation $\hat{W}_T$. Then we have computed numerically the highest eigenvalue
$\Lambda\{\hat{W}_T\}$ and extracted the braiding exponent by taking a limit $u\to\infty$ as
indicated in \eq{eq:5}. The results are summarized in \fig{fig:5} where the KPZ scaling is clearly
seen.

Few words should be said about the procedure of numerical computation of the resulting matrix
$\hat{W}_T$ and the extraction of the highest eigenvalue $\Lambda\{\hat{W}_T\}$. The matrix
multiplication is performed exactly in a symbolic form and each matrix element is the polynomial of
some degree of $u$. Then we have left in each matrix element, $W_{ij}$, only the monomial of
highest degree of $u$. Since later we have taken the limit $u\to\infty$ only such monomials
dominate. By Gershgorin theorem \cite{gersh} we have found the disk of radius $r<r(j)=\sum_{i=1}^n
|W_{ij}|$, where the highest eigenvalue is located and we have localized the position of this
highest eigenvalue. We have also checked each matrix $\hat{W}_T$ on the existence of zero's rows.
No such rows have been detected at least for $T<2\times 10^6$ and $n<400$.

The model described here can be extended to the case of a "symmetric random stirring protocol",
when the symmetric random walk on the full braid group $B_{n+1}$ is considered. For this situation
we expect the same KPZ scaling for the variance of the braiding exponent, however the description
of this process is much more involved because the random walk on the full braid group is not a
Markovian process. Namely, the cancellation in the word two sequentially added opposite generators
at times $t$ and $t+1$, like $\sigma_i(t)\, \sigma_i^{-1}(t+1) = \hat{e}$ demands the knowledge of
the full prehistory since the initial time moment $t=0$. This makes the exact analytic approaches
to the random walks on the full braid group rather doubtful. Nevertheless, we hope that some
approximative methods like the one developed in \cite{nech} should still be valid and can be used
for the investigation of random mixing.

The authors are grateful to S. Majumdar for valuable discussions.

\end{document}